\documentclass{article}
\usepackage{spconf,amsmath,graphicx,hyperref}
\usepackage[capitalize]{cleveref}
\usepackage{booktabs}

\title{Multi-bit Audio Watermarking}
\name{Luca A. Lanzendörfer \qquad Kyle Fearne \qquad Florian Grötschla \qquad Roger Wattenhofer}
\address{ETH Zurich}
\begin{document}
\maketitle
\begin{abstract}
We present Timbru, a post-hoc audio watermarking model that achieves state-of-the-art robustness and imperceptibility trade-offs without training an embedder-detector model. Given any 44.1 kHz stereo music snippet, our method performs per-audio gradient optimization to add imperceptible perturbations in the latent space of a pretrained audio VAE, guided by a combined message and perceptual loss. The watermark can then be extracted using a pretrained CLAP model. We evaluate 16-bit watermarking on MUSDB18-HQ against AudioSeal, WavMark, and SilentCipher across common filtering, noise, compression, resampling, cropping, and regeneration attacks. Our approach attains the best average bit error rates, while preserving perceptual quality, demonstrating an efficient, dataset-free path to imperceptible audio watermarking.
\end{abstract}
\section{Introduction}
\label{sec:intro}

Audio watermarking embeds imperceptible, machine verifiable signals into audio to support provenance, attribution, and copyright protection. This capability is increasingly critical in the era of social media and rapidly improving generative models, which enable the production and dissemination of highly realistic synthetic audio. Reliable watermarking can help end-users verify the legitimacy of clips, deter unauthorized sampling, and credit creators, while simultaneously raising the stakes for adversaries who seek to remove or forge watermarks.

Historically, audio watermarking was largely based on empirical schemes such as Quantization Index Modulation~\cite{QIM}, patchwork algorithms~\cite{patchwork}, least significant bit embedding~\cite{LSB}, and spread-spectrum techniques~\cite{SSM}. Although effective in certain settings, these methods often fail under common transformations such as audio compression. The trade-off between watermark imperceptibility and robustness against attacks remains at the center of audio watermarking and motivates our work. 

Recent learning-based approaches have made significant progress, spanning passive detectors~\cite{passive1, passive2} and joint embedded-detector architectures~\cite{pavlovic, dear, AudioSeal, WavMark, silentcipher} trained end-to-end. Passive detection is becoming increasingly less effective due to high-fidelity synthetic audio that closely mimics genuine content. In general, current watermarking approaches can be further categorized into ad-hoc and post-hoc methods. Ad-hoc models integrate watermarking within a generator to emit user- or model-specific watermarks~\cite{san2025latent}; post-hoc methods watermark arbitrary inputs after the fact. The latter offers greater flexibility and accessibility, enabling users to protect existing and novel content alike. Examples of recent post-hoc watermarking methods which jointly train an embedder and a detector include Wavmark~\cite{WavMark}, AudioSeal~\cite{AudioSeal} and SilentCipher~\cite{silentcipher}. AudioSeal proposes watermark detection at a sample level allowing for robust detection. Wavmark introduces a brute-force detection algorithm that also embeds a detection string before the payload in order to address issues with watermark localization. These two methods allow for watermarking of 16 kHz mono-channel audio snippets. Since neither method supports native stereo watermarking, we embed a watermark per channell.
SilentCipher places an emphasis on imperceptibility, allowing for a lower-bound on the Signal-to-Distortion Ratio (SDR) to be enforced. It also expands previous work to allow for 44.1 kHz stereo audio to be watermarked. These different methods emphasize the trade-off that exists in this domain between robustness against attacks and watermark imperceptibility.

\begin{figure*}[t!]
    \centering    
    \includegraphics[width=\textwidth]{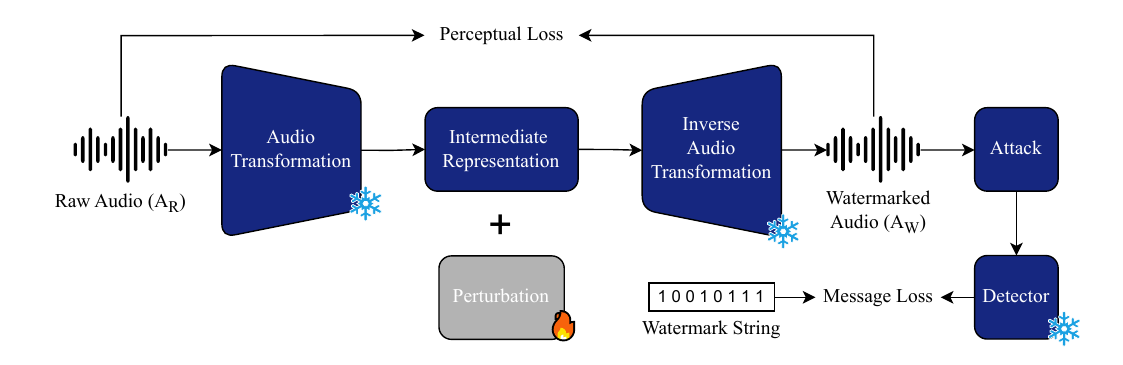}
    \caption{Overview of our proposed approach. The raw waveform $A_R$ is first transformed into the latent representation using a pretrained Stable Audio Open VAE. To embed a watermark, minor perturbations are added to this intermediate representation. At every step, this representation is decoded back into a waveform ($A_W$) and then augmented to simulate a variety of attacks. The perceptual loss and the message loss from the decoded message are then used to calculate the gradient which optimizes the perturbations. All other components remain frozen.}
    \label{fig:overview}
\end{figure*}

In this work, we propose Timbru, a post-hoc optimization-based method that performs gradient updates on a single stereo audio snippet, by perturbing the audio imperceptibly until a watermark is obtained that is robust to a wide range of attacks. This eliminates the compute and data requirements of training dedicated embedder-detector models and does not necessitate domain-specific fine-tuning for speech, music, or environmental audio.

Our contributions can be summarized as follows. We propose a post-hoc audio watermarking approach for 44.1 kHz stereo audio. Our approach encodes the audio using a pretrained Stable Audio Open VAE~\cite{oobleck}, which is then perturbed using gradient optimization to obtain an imperceptible watermark. To detect a watermark and its payload, we use a pretrained CLAP~\cite{CLAP} model as the feature extractor for watermark detection. We find that our approach is on average more robust to attacks while achieving similar perceptual quality compared to previous state-of-the-art methods.

\section{Methodology}
\label{sec:pagestyle}

The core idea behind Timbru is that perturbations are added to a latent representation of the audio during the optimization process in order to embed a watermark string of $k$ bits $m = m_1,...,m_k$ into the audio snippet, as shown in \cref{fig:overview}. The purpose of such perturbations is to modify the audio's features in a way that aligns with a secret key held by the user~\cite{freqmark,ssl}.
In a multi-bit setting, each user has a \textit{secret key} consisting of $k$ randomly selected orthogonal vectors. Each vector $v_1,...,v_k$ corresponds to an encoded bit. During the optimization process, the message $m$ is modulated into the signs of the projection of the features extracted by a pretrained CLAP model, $\phi(A_W)$, against each of the carriers.  The detector component then retrieves $\tilde{m}$ as follows:
\begin{equation}
     \tilde{m} = [sign(\phi(A_W)^\top v_1),...,  sign(\phi(A_W)^\top v_k)]
\end{equation}
 
\noindent \textbf{Training Pipeline.}
Audio waveform snippets $A_R$ are passed through a transformation stage $T(\cdot)$ in order to extract an embedding space within which to embed the watermark. 
We define $T(\cdot)$ to be passing the waveform through the Stable Audio Open VAE~\cite{oobleck} such that the intermediary representation can be written as \begin{equation}
    A_I = T(A_R) = Enc(A_R) 
\end{equation}
Small perturbations $\delta_m$ are then added to the intermediary representation $A_I$ and the inverse transformation is applied to convert the latent back to a raw audio waveform such that the resultant watermarked audio is:
\begin{equation}
    A_W = T^{-1}(A_I + \delta_m) = Dec(A_I+\delta_m)
\end{equation}
During the optimization stage, before detecting the watermark in the audio snippet, the watermarked audio $A_W$ is subjected to a random attack to introduce robustness.
The attacked audio is then passed through a detector and the message is retrieved. The loss, composed of both the perceptual loss between $A_R$ and $A_W$ and the message loss is then calculated and the gradient propagated back to $A_I$, which acts as a perturbation $\delta_w$ added inside the latent space.

\smallskip
\noindent \textbf{Losses.}
To capture robustness and the ability to detect and decode a watermark, we use a message loss~\cite{ssl}.
The optimization objective is to align the audio features $x$ as closely as possible to the $k$ vectors $v_1,...,v_k$ that correspond to the encoded message. The message loss is a hinge loss with margin $\mu > 0$ on the projections, defined as
\begin{equation}
    L_m(A_W) = \frac{1}{K} \sum_{k=1}^{K} max(0,\mu - (x^\top v_i).m_i),
\end{equation}
where $m = (m_1,...m_k) \in \{-1,1\}_k$ is the hidden message we embed in the audio snippet. The margin is set to $\mu=5$.

Additionally, a perceptual loss is used to ensure that any perturbations added to the audio remain imperceptible to humans. This perceptual loss, $L_p$, is taken from DAC~\cite{DAC} and consists of a combination of different losses, including a multi-scale Mel Spectrogram loss, as well as an adversarial discriminator loss. The total loss is therefore
\begin{equation}
    L = \lambda_mL_m + \lambda_pL_p,
\end{equation}
where $\lambda_m = 160$ and $\lambda_p = 4$ were empirically chosen as the optimal message weight and perceptual weight, respectively.

\begin{table*}[t!]
\centering

\begin{tabular*}{\linewidth}{@{\extracolsep{\fill}}lcccccccccc@{}}
\toprule
\textbf{Model} & \textbf{None} & \textbf{BP} & \textbf{LP} & \textbf{HP} & \textbf{E} & \textbf{S} & \textbf{DA} & \textbf{BA} & \textbf{GN} & \textbf{PN} \\
\midrule
AS~\cite{AudioSeal} & 1.58 & \textbf{1.75} & \textbf{41.00} & 61.13 & \textbf{2.63} & 5.25 & 1.58 & 1.54 & \textbf{9.54} & 1.63 \\
WM~\cite{WavMark}  & 0.55 & 2.58 & 49.92 & \textbf{0.64} & 14.75 & 4.16 & 0.55 & 0.54 & 48.90 & 0.95 \\ 
SC~\cite{silentcipher} & \textbf{0.01} & 23.59 & 48.84 & 4.36 & 11.32 & 8.56 & \textbf{0.01} & \textbf{0.01} & 50.88 & \textbf{0.38}\\
Timbru               & 0.83 & 17.5 & 53.30 & 25.00 & 22.5 & \textbf{0.00} & 0.83 & 0.42 & 20.42 & 2.5 \\
\bottomrule
\end{tabular*}

\vspace{0.6em} %

\begin{tabular*}{\linewidth}{@{\extracolsep{\fill}}lcccccccccc@{}}
\toprule
\textbf{Model} & \textbf{MP3} & \textbf{AAC} & \textbf{RS} & \textbf{Q} & \textbf{SS} & \textbf{RC} & \textbf{Speed} & \textbf{EnC.} & \textbf{Regen.} & \textbf{Avg.} \\
\midrule
AS~\cite{AudioSeal} & \textbf{1.79} & 42.83 & 1.58 & 1.75 & \textbf{2.50} & 42.92 & 43.83 & \textbf{6.96} & 66.46 & 17.79 \\
WM~\cite{WavMark}  & 11.05 & \textbf{10.44} & 0.55 & 1.23 & 32.35 & 43.22 & 50.30 & 49.37 & 49.24 & 19.54 \\
SC~\cite{silentcipher} & 37.46 & 37.86 & \textbf{0.01} & \textbf{0.44} & 50.46 & 37.75 & 49.50 & 50.08 & 49.39 & 24.25 \\
Timbru               & 5.42 & 22.08 & 0.83 & 1.67 & 6.67 & \textbf{30.83} & \textbf{40.00} & 10.41 & \textbf{21.67} & \textbf{14.89} \\
\bottomrule
\end{tabular*}
\caption{Results for 16-bit watermarking. We compare Timbru against AudioSeal (AS), WavMark (WM), and SilentCipher (SC) in terms of bit error rate (lower is better). We evaluate the watermarking models on bandpass (BP), lowpass (LP), highpass (HP), echo (E), smoothing (S), duck audio (DA), boost audio (BA), gaussian noise (GN), pink noise (PN), resampling (RS), quantization (Q), sample suppression (SS), random cropping (RC), EnCodec re-encoding (EnC.), and regeneration attack (Regen.). More details on the attack parameters used can be found in Section~\ref{sec:Results}. Whilst each method demonstrates their own clear advantages and disadvantages, on average, our method demonstrates the best average bit error rate, and notably outperforms previous methods on unseen regeneration attacks.}
\label{tab:results_multibit}
\end{table*}

\section{Experiments \& Results}
\label{sec:Results}

\begin{table}[t!]
\centering
\begin{tabular}{lrrrr}
\toprule
\textbf{} & \textbf{ViSQOL} $\uparrow$ &\textbf{SI-SNR (dB)} $\uparrow$  & \textbf{MUSHRA} $\uparrow$ \\
\midrule
AS~\cite{AudioSeal} & 1.91$\pm$0.54 & 19.65$\pm$6.18 & 57.18$\pm$3.22\\
WM~\cite{WavMark} & 1.91$\pm$0.53 & \underline{23.03$\pm$5.16} & 58.52$\pm$3.30 \\
SC~\cite{silentcipher}& \textbf{4.39$\pm$0.17}&\textbf{25.59$\pm$1.94} & \textbf{86.35$\pm$2.33} \\
Timbru & \underline{4.08$\pm$0.25} & 5.15$\pm$3.13 & \underline{66.32$\pm$3.52} \\
\bottomrule
\end{tabular}
\caption{Results for perceptual audio quality for 16-bit watermarking. We evaluate perceptual audio quality on ViSQOL, SI-SNR, and by conducting a MUSHRA human evaluation study. For ViSQOL and SI-SNR we show the standard deviation and for MUSHRA the 95\% confidence interval. We find that while SilentCipher achieves the best perceptual scores thanks to its SDR-bounded output, Timbru's perceptual quality is comparable while achieving higher detection accuracies.}
\label{tab:audio_multibit}
\end{table}

In line with previous work~\cite{WavMark,AudioSeal}, we embed 16 bits as our watermark message payload. We randomly pick 10\% of MUSDB18-HQ~\cite{musdb18-hq} mixtures and crop out 10-second snippets of these samples to evaluate the methods. We test the robustness of our approach against a variety of attacks using bit error rate metric to measure watermark message retrieval accuracy. In cases where decoding fails, a BER of 0.5 is assumed. In addition, we use ViSQOL~\cite{visqol} and SI-SNR~\cite{luo2018tasnet} to measure objective perceptual quality, as well as conducting a MUSHRA~\cite{MUSHRA} human evaluation study with 40 participants,\footnote{MUSHRA was conducted on \url{https://www.mabyduck.com}} where each participant was asked to score watermarked audio on perceptual quality. The subjective perceptual study contained one hidden reference, and two anchors (3.5 kHz, 7 kHz) as well as four stimuli (Timbru, WavMark, SilentCipher and AudioSeal). The participants were briefed beforehand about the task and were asked to rate the perceptual quality of each stimuli. Each participant first listened and ranked two practice trials, which were randomly sampled from the 15 samples, and then completed five trials. Each trial consisted of a random sample and participants took a mandatory short break between trials.
\begin{figure*}[!htbp]
    \centering    
    \includegraphics[width=\textwidth]{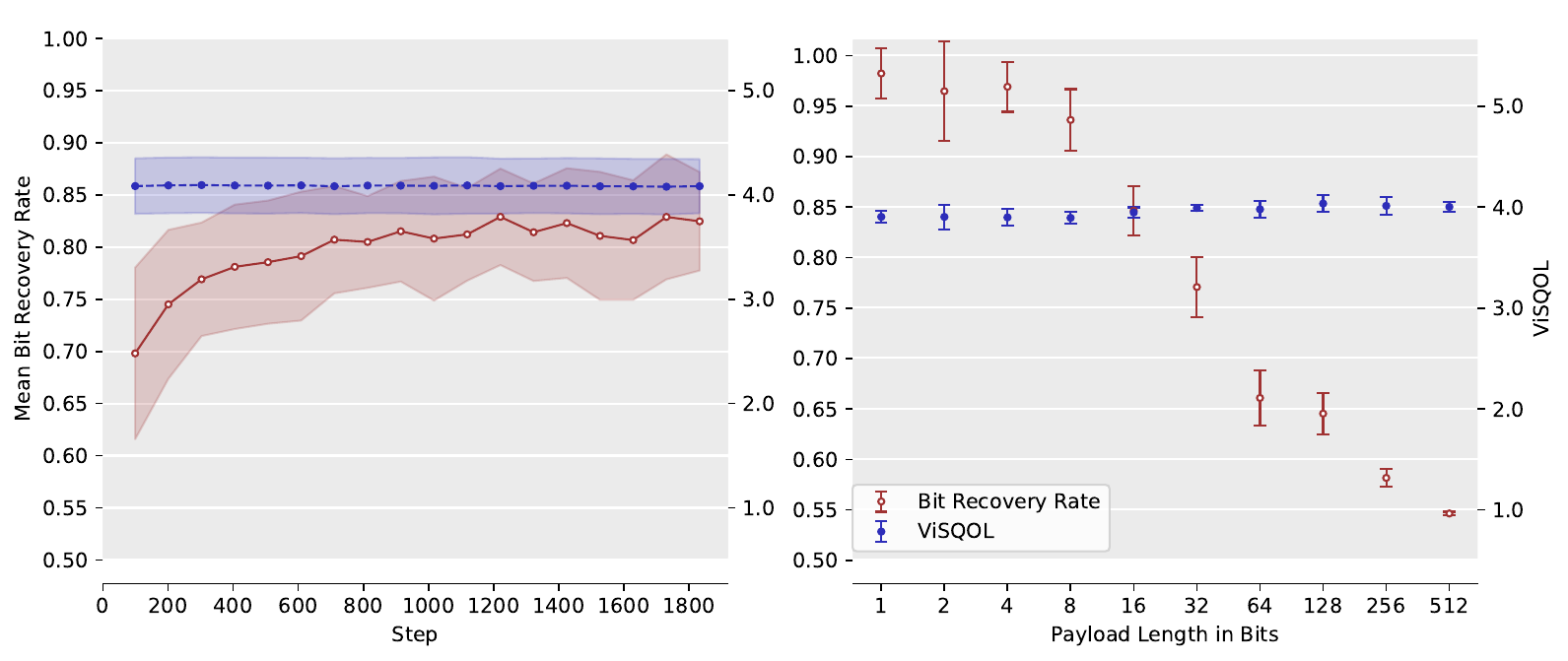}
    \caption{(Left) Mean bit recovery rate ($\mathrm{BRR}=(1-\mathrm{BER})/100$) for 16-bit payload over optimization steps shows the longer we run Timbru, the more robust the embedded watermark becomes. (Right) Ablation where each point represents the mean BRR for watermarked audio with specific payload length, showing how the mean BRR and the perceptual quality change as the payload length increases.}
    \label{fig:weighting_plot}
\end{figure*}

\smallskip
\noindent \textbf{Attack parameters.} Timbru, AudioSeal~\cite{AudioSeal}, WavMark~\cite{WavMark}, and SilentCipher~\cite{silentcipher} were evaluated against a variety of attacks which could potentially be used as means for watermark removal (inadvertently or through malicious intent). The attacks are common among other audio watermarking methods~\cite{AudioSeal, WavMark, silentcipher} and the attacks were performed using the Audiocraft library~\cite{audiocraft}. The parameters were chosen to reflect the evaluation from AudioSeal~\cite{AudioSeal}. First, the watermarking methods were evaluated against a set of spectral filtering attacks, namely a bandpass filter (500–5000 Hz), low-pass filter (cut-off 500 Hz), high-pass filter (cut-off 1500 Hz) and smoothing with a moving-average window of 40 samples. Attacks modulating the amplitude were also carried out, such as ducking (gain of 10), boosting (gain of 0.1) and sample suppression (3\% of samples set to zero). Another set of attacks dealt with temporal alterations, involving random cropping to 50\% of the original duration, an echo with 0.5s delay at 0.5 relative volume and a speed change at a factor of 1.25. Attacks that introduced sampling artifacts were also evaluated, such as quantization to $2^9$ levels and resampling to 32kHz. We also tested compression with lossy codecs and a neural audio codec: AAC compression at 64 kbps, MP3 compression at 32 kbps, and EnCodec~\cite{Encodec} by re-encoding at 24kHz and then resampling back to 44.1kHz. Furthermore, additive noise attacks were used, including pink noise ($\sigma=0.1$) and Gaussian noise ($\sigma=0.05$). Finally, we also evaluated against a strong regeneration attack which was not seen during training, and involved re-encoding audio using the DAC~\cite{DAC} model at 44.1kHz. During Timbru's training, the parameters for these attacks were sampled randomly from a weaker parameter range than what we evaluated against. Non-differentiable attacks such as MP3 and AAC compression are implemented using a straight-through estimator~\cite{straight-through} to allow back-propagation of the gradients.

The bit error rates for each attack are shown in \cref{tab:results_multibit}. While our method achieves higher average message-reconstruction accuracy than AudioSeal~\cite{AudioSeal}, WavMark~\cite{WavMark}, and SilentCipher~\cite{silentcipher}, each watermarking approach has distinct strengths and weaknesses. Audioseal proves to be robust against sample suppression due to its sample-level localization techniques that implement sample-level masking during training. Bandpass and Lowpass results show that AudioSeal also demonstrates its strength by not encoding a watermark in the low- or high-frequency domain, unlike WavMark, which tends to encode its watermarks in the high frequencies. SilentCipher outperforms all other approaches in terms of imperceptibility through its use of a signal distortion bound, however, this also causes it to suffer the most in terms of bit recovery rate. Furthermore, it is interesting to note that, compared to other methods, Timbru offers the best robustness against unseen regeneration attacks, which tend to be the most difficult attack type to defend against. 
Since we use CLAP to extract features that are used to detect the watermark, and that CLAP extracts features from Mel, it is likely that the watermark is visible in the Mel Spectrogram. Therefore, we believe that this is the reason why the regeneration attack is not as effective compared to other watermarking approaches.

Analyzing the watermarked audio quality in \cref{tab:audio_multibit}, we find that SilentCipher~\cite{silentcipher} offers better general audio quality as measured by the objective metrics and by the participants in the MUSHRA listening study. Thanks to its distortion-bound this is not entirely surprising. For the MUSHRA study, the participants rated the reference, mid-anchor (7 kHz), and low anchor (3.5 kHz) as 89.16$\pm$2.19, 52.57$\pm$3.46, and 17.23$\pm$2.53, respectively. The significantly lower performance of Timbru in terms of SI-SNR can be explained due to the audio being passed through a VAE, which can cause a variety of signal-level artifacts that are imperceptible to humans (e.g., sample mismatch, phase inversion).
In \cref{fig:weighting_plot} we show the performance of Timbru in terms of optimization steps. Unsurprisingly, we find that the longer we optimize, the more robust the watermark becomes, although there are diminishing returns after a few thousand steps. For our experiments, we set a stopping condition if the bit recovery rate does not improve for 1k steps. On average, the watermarking process takes roughly one hour per audio snippet.
Furthermore, we show the trade-off between the number of bits in the payload and the corresponding ViSQOL score. We find that as the number of bits increases, the robustness against attacks tends to degrade.

\smallskip
\noindent \textbf{Conclusion.}
We introduced Timbru, a post-hoc audio watermarking method that preserves perceptual quality while improving robustness by performing per-snippet gradient optimization to embed small perturbations in a latent representation of audio, offering a strong dataset-free alternative to state-of-the-art watermarking approaches.

\vfill\pagebreak

\ninept
\bibliographystyle{IEEEbib}
\bibliography{strings,refs}

\end{document}